\documentclass[12pt]{article}

\usepackage[dvips]{graphicx}
\usepackage{epsfig}
\usepackage{amsmath,amsfonts,amssymb,amsthm}
\usepackage{mathrsfs,mathtools}
\usepackage{verbatim}
\usepackage{psfrag}
\usepackage{bm}
\usepackage[english]{babel}
\usepackage{bbm}
\usepackage[utf8]{inputenc}
\usepackage[square,comma,sort&compress,numbers]{natbib}
\usepackage[dvipsnames]{xcolor}
\usepackage{slashed}
\usepackage{upgreek}
\usepackage{enumitem}
\usepackage{float}
\usepackage[T1]{fontenc}
\usepackage{soul}
\usepackage{pgfplots}
\usepackage{extarrows}
\usepackage{appendix}
\pgfplotsset{compat=1.18}
\usetikzlibrary{external}

\usepackage[T1]{fontenc}

\usepackage[colorlinks=true,urlcolor=blue,anchorcolor=blue,citecolor=blue,filecolor=blue
,linkcolor=blue,menucolor=blue,linktocpage=true,pdfproducer=medialab,pdfa=true
]{hyperref}
\usepackage{epsf,epsfig}
\usepackage{graphics}
\usepackage{subcaption}
\usepackage{moresize}
\usepackage{cancel}

\newcommand{\cb}{\color{black}}
\newcommand{\cbl}{\color{black}}
\newcommand{\bL}{\begin{Large}}
\newcommand{\eL}{\end{Large}}
\newcommand\blfootnote[1]{%
  \begingroup
  \renewcommand\thefootnote{}\footnote{#1}%
  \addtocounter{footnote}{-1}%
  \endgroup
}

\def\d{\mathrm{d}}

\def\vec{\mathbf}

\def\e{\mathrm{e}}

\newcommand{\cPT}{\ensuremath{\mathcal{PT}}}

\newcommand{\bea}{\begin{eqnarray}}
\newcommand{\eea}{\end{eqnarray}}
\newcommand{\be}{\begin{equation}}
\newcommand{\ee}{\end{equation}}
\newcommand{\bal}{\begin{align}}
\newcommand{\eal}{\end{align}}

\usepackage{tikz} 
\usetikzlibrary{shapes.misc}

\usepackage[errorstop]{feynmp}
\newcommand{\rom}[1]{\uppercase\expandafter{\romannumeral #1\relax}}

\begin{document}

\thispagestyle{empty}

\begin{flushright}
{\small KCL-PH-TH/2024-32}
\end{flushright}

\vspace{0.4cm}

\begin{center}
\Large\bf\boldmath
Piecewise linear potentials for false vacuum decay and negative modes 
\end{center}

\vspace{-0.2cm}

\begin{center}
{Wen-Yuan Ai,*\blfootnote{*~wenyuan.ai@kcl.ac.uk} Jean Alexandre$^\dag$\blfootnote{$^\dag$~jean.alexandre@kcl.ac.uk} and Sarben Sarkar$^\ddag$\blfootnote{$^\ddag$~sarben.sarkar@kcl.ac.uk} \\
\vskip0.4cm
{\it Theoretical Particle Physics and Cosmology, King’s College London,\\ Strand, London WC2R 2LS, UK}\\
\vskip1.4cm}
\end{center}

\begin{abstract}
We study bounce solutions and associated negative modes in the class of piecewise linear triangular-shaped potentials that may be viewed as approximations of smooth potentials. In these simple potentials, the bounce solution and action can be obtained analytically for a general spacetime dimension $D$. The eigenequations for the fluctuations around the bounce are universal and have the form of a Schr\"odinger-like equation with delta-function potentials. This Schr\"odinger equation is solved exactly for the negative modes whose number is confirmed to be one.
The latter result may justify the usefulness of such piecewise linear potentials in the study of false vacuum decay.
\end{abstract}

\newpage

\hrule
\tableofcontents
\vskip.85cm
\hrule

\section{Introduction }
\label{sec:Introduction}

In the semiclassical analysis of false vacuum decay~\cite{Coleman:1977py,Callan:1977pt,Coleman:1985rnk,Langer:1969bc,Bender:2023cso} based on Euclidean path integrals, a special classical solution to the Euclidean equation of motion, called the bounce, plays a key role. The decay rate $\Gamma$ of the false vacuum generally takes the form of
\be
\label{eq:decay_rate}
\Gamma =A\exp \left( -B/\hbar \right)  \left( 1+O\left( \hbar \right)  \right)\,  
\ee
where $A$ and $B$ are potential-dependent constants.
At the Gaussian level (taking $\hbar=1$), Callan and Coleman derived the decay formula per unit volume as ~\cite{Callan:1977pt}
\begin{align}
\label{eq:decay-oneloop}
\frac{\Gamma}{V}=\left(\frac{B}{2\pi}\right)^{D/2}\left|\frac{\det^\prime[-\partial^2+U''(\phi_b)]}{\det[-\partial^2+U''(\phi_{+})]}\right|^{-1/2}\,\e^{-B}\,,
\end{align}
where $\phi_{+}$ and $\phi_b$ are the false vacuum and the classical bounce, respectively, $B\equiv S[\phi_b]-S[\phi_+]$ is the normalized bounce action (given in terms of the classical action $S$), and $D$ is the spacetime dimension. In Eq.~\eqref{eq:decay-oneloop}
$\det'$ means that the zero
eigenvalues are omitted from the determinant and a prefactor $\sqrt{B/2\pi}$ is included for each of the $D$ collective coordinates that are related to the zero modes corresponding to spacetime translations~\cite{Gervais:1974dc}.

The Callan-Coleman formula for the decay rate is derived from the imaginary part of the false-vacuum energy through $\Gamma=-2{\rm Im}E_0$. For the imaginary part of the ground-state energy to be nonvanishing, $\det^\prime[-\partial^2+U''(\phi_b)]$ needs to be negative. The imaginary part ${\rm Im}E_0$ is then obtained through the so-called ``potential deformation'' method~\cite{Callan:1977pt}.\footnote{See Refs.~\cite{Andreassen:2016cvx,Ai:2019fri} for a more rigorous explanation using the Picard-Lefschetz theory~\cite{pham1983vanishing,berry1991hyperasymptotics,Witten:2010cx,Witten:2010zr}.}  For this derivation to be valid, the number of the negative eigenfunctions in the spectrum of the fluctuation operator $-\partial^2+U''(\phi_b)$ must be an odd number, and typically it is believed to be just one, although there is no proof of this in field theory. 
 
In the case of a finite set of coupled oscillators, Coleman gave an intuitive argument \cb which leads to a conjecture \cbl that there is only one negative eigenvalue~\cite{Coleman:1987rm} in the absence of gravitational effects.\footnote{In the presence of gravity there have been cases where the spectrum for bounces has more than one negative eigenvalues~\cite{Battarra:2012vu,Gregory:2018bdt}.} \cb At a \emph{formal} level a scalar field theory may be represented as the limiting behaviour of an infinite number of coupled oscillators. However a careful argument \cite{Glimm1987-pw,Glimm:1979zp,PhysRevD.8.3346} shows that a suitable  finite set of quartically coupled oscillators is equivalent to the two dimensional scalar field theory $\left( \phi^{4} \right)_{2}$ with ultraviolet and infrared cutoffs. \cbl Hence the arguments of Coleman are not immediately applicable in a field theoretic context, \cb as stated in \cite{Coleman:1987rm}, \cbl and part of our purpose is to investigate the validity of Coleman's conjecture in field theory within the class of piecewise linear potentials. \cb In Appendix~\ref{app:Colemann}, we review Coleman's arguments, which intuitively show that there should be one and only one negative mode for the bounce solutions. \cbl

In field theory, the determination of the number of negative eigenvalues cannot be performed rigorously for general potentials since the bounce solutions are not known analytically, except in very restricted cases~\cite{Fubini:1976jm,Lipatov:1976ny,FerrazdeCamargo:1982sk,Coleman:1985rnk,Lee:1985uv,Andrianov:2011fg,Aravind:2014pva,Bender:2023cso} and for some of these the prefactor $A$ may be analytically computed at the Gaussian level~\cite{Garbrecht:2018rqx,Guada:2020ihz}. In general, only numerical bounce solutions can be obtained.\footnote{In recent years, a new formalism, called the tunnelling potential method, has been developed to calculate analytically the tunnelling action~\cite{Espinosa:2018hue} that can be applied for cases with gravity~\cite{Espinosa:2021tgx} and multiple fields~\cite{Espinosa:2023oml}.} This makes analysing the eigenvalue equations for the fluctuations around the bounce difficult. In many cases, the number of the negative modes can be numerically confirmed to be odd by using the Gel'fand-Yaglom method~\cite{Gelfand:1959nq} (see e.g. Refs.~\cite{Dunne:2005rt,Ai:2023yce}). Even with only a numerical bounce solution, the eigenvalue problem (including the computation of negative eigenvalues) can, in principle, be handled numerically upon a discretization of the fluctuation operator~\cite{Ekstedt:2023sqc}.

For scalar field theories, one way of generating potentials, which have explicit closed-form bounce-like solutions, is to consider potentials from piecewise linear functions (also referred to as  polygonal potentials in general)~\cite{Duncan:1992ai,Guada:2018jek}. Clearly, any smooth potential graph can be suitably sampled at a discrete set of points, which can be sequentially connected by straight lines to give a polygonal graph. Some simple polygonal potentials were proposed in Refs.~\cite{Duncan:1992ai,Dutta:2011rc,Dutta:2012qt,Guada:2018jek}.
In this work, we focus on the triangular potential~\cite{Duncan:1992ai} and the Lee-Weinberg potential~\cite{Lee:1985uv}. The latter is unbounded from below and could be used as a model for non-Hermitian $\cPT$-symmetric theory~\cite{R2,Ai:2022csx}.  Smooth, exact analytic bounce solutions can be found for these potentials. In addition, there is a crucial and simple feature of such polygonal potentials; $U''(\phi)$ is nonvanishing only at some finite points in the field space, making the eigenequation extremely simple and universal. Hence such piecewise linear potentials allow an analytic study of the negative eigenmodes associated with the bounce for arbitrary $D$.

The outline of the paper is as follows. In Sec.~\ref{sec:models}, we introduce the two types of potentials mentioned above and derive the analytic bounce solutions and actions for general spacetime dimensions. In Sec.~\ref{sec:eigenequations} we study the eigenequations for the fluctuation operator evaluated at the bounce, focusing on the number of negative modes. We find that for all the cases that we study, there is one and only one negative mode in the spectrum. This may justify Coleman's conjecture \cite{Coleman:1987rm} in a field theory setup for any spacetime dimensions.  We conclude in Sec.~\ref{sec:conclusion}.

\section{Polygonal-potential models for false vacuum decay and classical bounces}
\label{sec:models}

\subsection{Lee-Weinberg potential}

\begin{figure}[ht]
    \centering
    \includegraphics[scale=0.7]{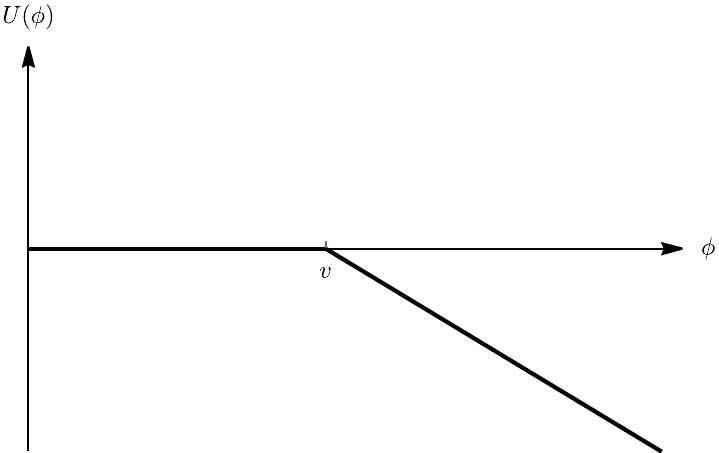}
    \caption{The Lee-Weinberg potential studied in Ref.~\cite{Lee:1985uv}.}
    \label{fig:LW-pot}
\end{figure}

We first consider a potential introduced in Ref.~\cite{Lee:1985uv}, which we refer to as the Lee-Weinberg potential below, reading
\begin{align}
    U(\phi)=\begin{cases}
   0\,, \qquad &{\rm for \ }|\phi|<v\,, \\
    -K\, (|\phi|-v)\,, \qquad &{\rm for\ } |\phi|>v\,.
    \end{cases}
\end{align}
This gives, for $\phi\geq 0$,
\begin{align}
    U'(\phi)=-K\Theta(\phi- v)\,,\qquad {\rm and}\qquad U''(\phi)=-K\delta(\phi-v)\,,
\end{align}
where $\Theta(x)$ is the Heaviside step function. The potential is shown in Fig.~\ref{fig:LW-pot}.

Even though there is no potential barrier in the Lee-Weinberg potential, there are still bounce solutions as long as $D\geq 3$. These bounce solutions describe ``tunnelling without barrier''. Mathematically, a bounce solution is possible because of the ``friction term'' in the equation of motion (see below). This is reminiscent of the famous Fubini-Lipatov instanton \cite{Fubini:1976jm,Lipatov:1976ny} for the $-\lambda\phi^4$ (with $\lambda>0$) in four-dimensional spacetime, which is also used as a non-Hermitian $\cPT$-symmetric model~\cite{Bender:1998ke,Ai:2022csx}. Below, we consider the tunnelling from $\phi=0$ to a point $\phi_*>v$ where the exit point $\phi_*$ is determined by the solution. The bounce solution has $O(D)$ symmetry and satisfy
\begin{align}
\label{eq:bounce-EoM}
    -\frac{\d^2 \phi(r)}{\d r^2}-\frac{D-1}{r}\frac{\d\phi(r)}{\d r}+U'(\phi(r))=0\,,\quad \left.\frac{\d\phi}{\d r}\right|_{r=0}=0\,,\quad  \phi(r\rightarrow\infty)=0\,,
\end{align}
where $r=\sqrt{x^{2}_{1}+x^{2}_{2}+\ldots +x^{2}_{D}}$. The second term is the ``friction term'' we referred to above.  For $D=2$, the above boundary conditions cannot be satisfied for the Lee-Weinberg potential and thus one obtains a bounce solution only for $D\geq 3$. We will return to the case of $D=2$ when we discuss the triangular potential.

\begin{figure}[ht]
    \centering
    \includegraphics[scale=0.7]{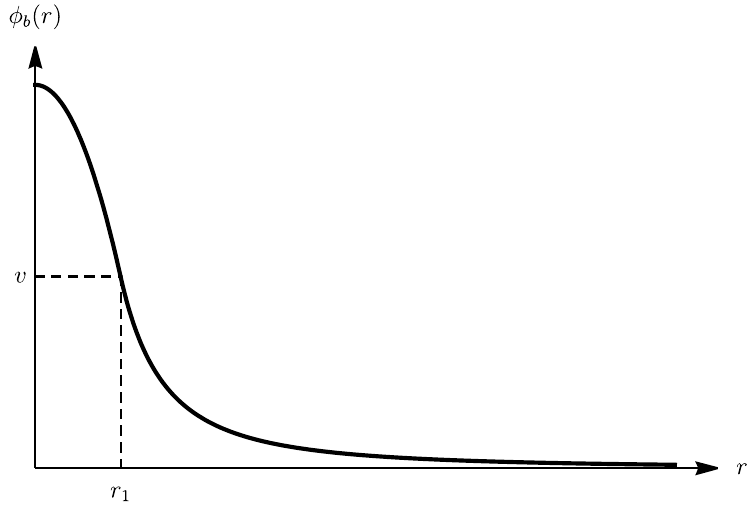}
    \caption{The bounce solution given in Eq.~\eqref{eq:bounce-LW} for $D=4$. }
    \label{fig:bounce}
\end{figure}

For $D\geq 3$ we find the solution as
\begin{align}
\label{eq:bounce-LW}
    \phi_b(r)=\Theta(r_1-r)
    v\left[\frac{D}{2}-\frac{(D-2)}{2}\frac{r^2}{r_1^2}\right] + \Theta(r-r_1)
    v\left(\frac{r_1}{r}\right)^{D-2} \,,
\end{align}
where 
\begin{align}
    r_1=\left(\frac{D(D-2)v}{K}\right)^{\frac{1}{2}}\,.
\end{align} 
One can read the exit point $\phi_*=\phi_b(0)=v D/2$.
The classical bounce action\footnote{The action $S_{\rm E}[\phi]$ is given by $$S_{\rm E}[\phi ]=\frac{2\pi^{D/2} }{\Gamma \left(\frac{D}{2} \right)} \int^{\infty }_{0} \d r\,  r^{D-1}\left( \frac{1}{2} \left( \frac{\d\phi}{\d r} \right)^{2}  +U\left( \phi \right)  \right)$$ with $\Gamma[x]$ being the gamma function.} is obtained as
\begin{align}
    B=\frac{4\pi^2(D-2)v^2}{D+2} \left(\frac{D(D-2)v}{K}\right)^{\frac{D-2}{2}}\,.
\end{align}
For $D=4$, we recover the results given in Ref.~\cite{Lee:1985uv}.
The functional determinant is 
\begin{align}
    A=\frac{\det'\left[-\partial^2+U''(\phi_b(r))\right]}{\det[-\partial^2] }= \frac{\det'\left[-\partial^2-\frac{D}{r_1}\delta(r-r_1)\right]}{\det[-\partial^2] }=r_1^{2D}\, \frac{\det' [-\partial^2-D\delta(r-1)]}{\det[-\partial^2]}\,,
\end{align}
where in the second step we have done the rescaling: $x\rightarrow r_1 x$, $\partial_x\rightarrow \partial_x/r_1$. The factor of $r_1^{2D}$ is a consequence of subtracting $D$ zero modes in the functional determinant in the numerator, which causes a mismatch in the total number of eigenmodes from the numerator and denominator. Taking $D=4$ as an example, we plot the bounce, which is smooth, in Fig.~\ref{fig:bounce}.

\subsection{Triangular potential}

\begin{figure}[ht]
    \centering
    \includegraphics[scale=0.8]{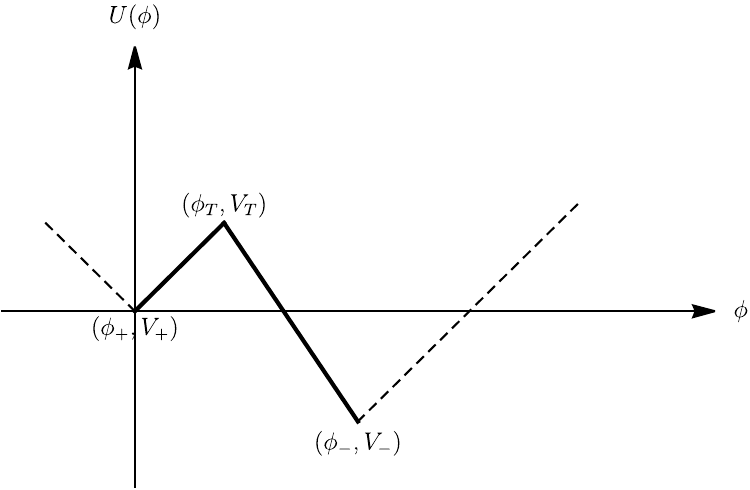}
    \caption{Triangular potential studied in Ref.~\cite{Duncan:1992ai}. We have taken $\phi_+=0$, $V_+=0$.}
    \label{fig:Tri-pot}
\end{figure}

Another simple model that can give exact bounce solutions is the triangular potential~\cite{Duncan:1992ai}, which (when the sharp bends in the potential are smoothed out) serves as a common paradigm for false vacuum decay. We illustrate this potential in Fig.~\ref{fig:Tri-pot}. The transition from the false vacuum $\phi_+$ to the true vacuum $\phi_-$ is sensitive only to the barrier part (thick lines in Fig.~\ref{fig:Tri-pot}) and therefore this model is characterized by the three pair of parameters $\{(\phi_{+}, V_{+}),(\phi_{T}, V_{T}),(\phi_{-}, V_{-})\}$. Without loss of generality, we take $\phi_+=0$, $V_+=0$. The potential thus reads
\begin{align}
\label{eq:U-Tri}
    U(\phi)=\begin{cases}
   \lambda_+ \phi\,, \qquad &{\rm for \ } 0<\phi<\phi_T\,, \\
    V_T-\lambda_-(\phi-\phi_T)\,, \qquad &{\rm for\ } \phi_T<\phi<\phi_-\,,
    \end{cases}
\end{align}
where
\begin{align}
    \lambda_+=\frac{V_T}{\phi_T}>0\,,\qquad \lambda_-=\frac{V_T-V_-}{\phi_--\phi_T}>0\,.
\end{align}
From the potential, one has (for $0\leq\phi\leq\phi_-$)
\begin{align}
    U'(\phi)=\lambda_+\Theta(\phi_T-\phi) -\lambda_-\Theta(\phi-\phi_T)\,,\quad {\rm and}\quad
    U''(\phi)=-(\lambda_+ + \lambda_-)\delta(\phi-\phi_T)\,.
\end{align}
For future use, we define the parameter 
\begin{align}
    c\equiv \frac{\lambda_-}{\lambda_+}\,.
\end{align}
Note that the Lee-Weinberg potential discussed in the last subsection can be viewed as a special case of this triangular potential, $\lambda_+=0$, $\lambda_-=K$, which gives $c\rightarrow \infty$. Although $c\rightarrow \infty$ is a physical limit, $c=0$ is a singular limit and is not physical because in that case, $\phi_+$ cannot be a false vacuum as we have assumed.

For $D=1$, the bounce solution is a function of the imaginary time $\tau\equiv x_1\in (-\infty,\infty)$ while for $D>1$, it is a function of the $D$-dimensional Euclidean radius $r\in [0,\infty)$. Below, we discuss $D=1$ and $D\geq 3$ cases separately, while leaving the case $D=2$ for Appendix~\ref{app:tri-D2}. 

\subsubsection{\texorpdfstring{$D=1$}{TEXT}}

\begin{figure}[ht]
    \centering
    \includegraphics[scale=0.8]{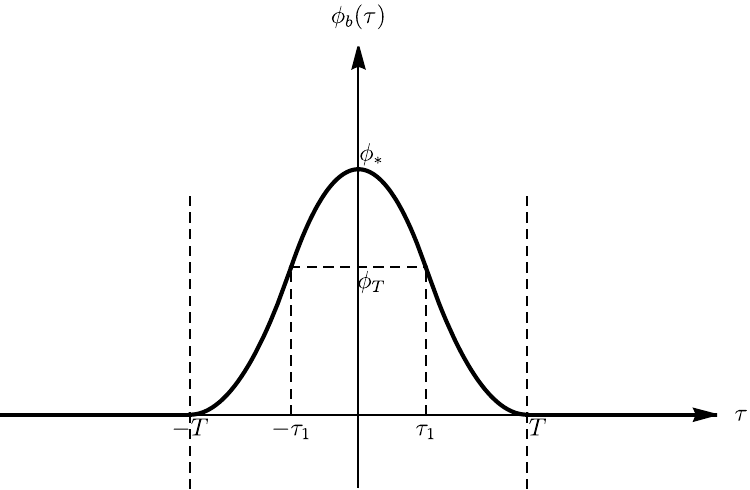}
    \caption{The bounce solution, Eq.~\eqref{eq:bounce-Tri}, for the triangular potential~\eqref{eq:U-Tri}. The bounce solution reaches the false vacuum at finite time $\pm T$. }
    \label{fig:bounce_Tri}
\end{figure}

For $D=1$, the bounce satisfies (taking the ``center'' of the bounce to be $\tau=0$)
\begin{align}
    -\frac{\d^2\phi(\tau)}{\d\tau^2}+U'(\phi(\tau))=0\,,\quad \left.\frac{\d\phi(r)}{\d\tau}\right|_{\tau=0}=0\,,\quad \phi(\tau\rightarrow\pm\infty)=0\,.
\end{align}
First, it is easy to check, by energy conservation in the Euclidean space, that the turning point $\phi_*$ is given by 
\begin{align}
    \phi_*=\phi_T+\frac{V_T}{\lambda_-}\,.
\end{align}
The bounce solution is found to be 
\begin{align}
\label{eq:bounce-Tri}
  \phi_b(\tau)=\begin{cases} 
     -\frac{1}{2}\lambda_- \tau^2+\phi_*\,, \qquad &{\rm for \ }|\tau|<\tau_1\,, \\
    \frac{1}{2}\lambda_+ (|\tau|-\tau_1)^2-\sqrt{2V_T} |\tau| + \phi_T+\frac{2V_T}{\lambda_-}\,, \qquad &{\rm for\ } \tau_1\leq |\tau|< T\,,\\
    0\,, \qquad &{\rm for \ }|\tau|\geq T\,,
    \end{cases}
\end{align}
where
\begin{align}
    \tau_1=\frac{\sqrt{2V_T}}{\lambda_-}\,, \qquad T=\tau_1+\frac{\sqrt{2V_T}}{\lambda_+}=\frac{\sqrt{2V_T}}{\lambda_-}+\frac{\sqrt{2V_T}}{\lambda_+}\,.
\end{align}
We show the bounce solution in Fig.~\ref{fig:bounce_Tri}.  The bounce action is found to be
\begin{align}
    B=\frac{4V_T\sqrt{2V_T}}{3}\left(\frac{1}{\lambda_+}+\frac{1}{\lambda_-}\right)\,.
\end{align}

Note that in the bounce solution, the field reaches the false vacuum $\phi_+=0$ at finite times $\pm T$. \cb This is because the potential is not smooth at the false vacuum. In Appendix~\ref{app:tri-mass} we study the case when the linear function around $\phi=0$ is replaced by a mass term. In that case, the field in the bounce solution does reach the false vacuum asymptotically at infinity.
\cbl

To analyse the eigenequations, it is sufficient for us to know the derivative of $\phi_b(\tau)$ at $\pm\tau_1$, $\phi_b(\tau)'|_{\tau=\pm\tau_1}=\mp\sqrt{2V_T}$. Therefore, the fluctuation operator at the bounce reads
\begin{align}
    -\partial_\tau^2+U''(\phi_b(\tau))=-\partial_\tau^2-\frac{(\lambda_+ +\lambda_-)}{\sqrt{2V_T}}\left[\delta(\tau+\tau_1)+\delta(\tau-\tau_1)\right]+ C \left[\delta(\tau+T)+\delta(\tau-T)\right]\,,
\end{align}
where $C$ is a positive parameter. The last two Dirac delta functions are due to the discontinuity of the potential at the false vacuum ($\phi_+=0$). These delta functions effectively restrict the range of $\tau$ to a finite region $[-T,T]$ and the specific form of $C$ is not needed for analysing the eigenequations. 
Thus the relevant fluctuation operator becomes 
\begin{align}
    -\partial_\tau^2-\frac{(\lambda_+ +\lambda_-)}{\sqrt{2V_T}}\left[\delta(\tau+\tau_1)+\delta(\tau-\tau_1)\right]\qquad {\rm with}\quad \tau\in[-T,T]\, .
\end{align}
To simplify further, we can rescale $\tau\rightarrow \tau_1 \tau$. Then the relevant fluctuation operator becomes
\begin{align}
   \frac{1}{\tau_1^2}\left( -\partial_\tau^2-\left(1+\frac{1}{c}\right)\left[\delta(\tau+1)+\delta(\tau-1)\right]\right)\qquad {\rm with}\quad \tau\in[-T/\tau_1,T/\tau_1]\, ,
\end{align}
where we recall $c=\lambda_-/\lambda_+$. The total factor $1/\tau_1^2$ does not affect the analysis of the number of negative modes.

\subsubsection{\texorpdfstring{$D\geq 3$}{TEXT}}

\begin{figure}[ht]
    \centering
    \includegraphics[scale=0.7]{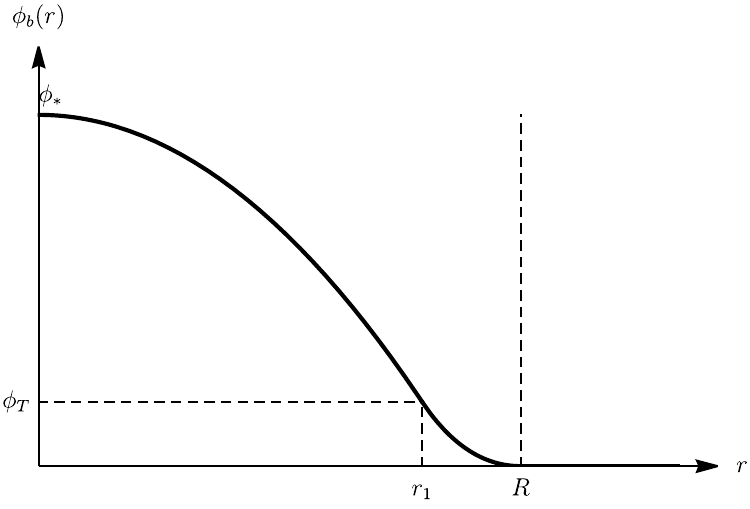}
    \caption{The bounce solution~\eqref{eq:bounce-tri-Dgeq3} for the triangular potential~\eqref{eq:U-Tri}. Independently of $D$, the bounce reaches the false vacuum at a finite radius $R$ because we have a nonvanishing slope at (or more precisely, infinitely close to) the false vacuum.}
    \label{fig:bounce-tri-Dgeq3}
\end{figure}

Now we consider $D\geq 3$. The general bounce solution that we find reads
\begin{align}
\label{eq:bounce-tri-Dgeq3}
    \phi_b(\tau)=\begin{cases} 
    -\frac{\lambda_-}{2D}r^2+\phi_*\,, \qquad &{\rm for \ } 0\leq r< r_1\,, \\
    \frac{\lambda_+}{2D}r^2+\frac{\lambda_+ R^D }{D(D-2)}r^{2-D}-\frac{\lambda_+ R^2}{2(D-2)}\,, \qquad &{\rm for\ } r_1\leq r < R\,,\\
    0\,, \qquad &{\rm for \ } r\geq R\,,
    \end{cases}
\end{align}
where 
\begin{subequations}
    \begin{align}
        &R=\left(\frac{2D(D-2)\phi_T (1+c)^{\frac{2}{D}}}{\lambda_+\left[D+2c-D (1+c)^{\frac{2}{D}}\right]}\right)^{\frac{1}{2}}\,,\\
        &\phi_*=\phi_T+\frac{c(D-2)\phi_T}{D+2c-D(1+c)^{\frac{2}{D}}}\,,\\
        &r_1=\left(\frac{2D(D-2)\phi_T}{\lambda_+\left[D+2c-D(1+c)^{\frac{2}{D}}\right]}\right)^{\frac{1}{2}}\,.
    \end{align}
\end{subequations}
The bounce action is 
\begin{align}
    B=\frac{4\pi^{\frac{D}{2}}}{\Gamma\left(\frac{D}{2}\right)}\frac{(1+c)\lambda_+\phi_T}{D(D+2)}r_1^D\,.
\end{align}
For $D=4$, we recover the results given in Ref.~\cite{Duncan:1992ai}. Taking $D=4$ as an example, we show the bounce solution in Fig.~\ref{fig:bounce-tri-Dgeq3}. Similar to the $D=1$ case, the bounce solution $\phi_b(r)$ reaches the false vacuum at finite $R$. 
And the delta-potentials at $r=\pm R$ in the fluctuation operator effectively restrict $r$ to $[0,R]$. The fluctuation operator at the bounce, after rescaling $r\rightarrow r_1 r$, reads
\begin{align}
\label{onedelta}
    \frac{1}{r_1^2}\left[-\partial^2- D\left(1+\frac{1}{c}\right)\delta(r-1)\right]\qquad {\rm with}\quad r\in[0,R/r_1],\quad D\geq 3\, .
\end{align}

\section{Eigenequations and negative modes}
\label{sec:eigenequations}

In this section, we study the eigenequation for the fluctuations about the bounce. We do not aim to solve the full eigenspectrum but focus on the negative eigenmodes. 
For the analytic bounce solutions to faithfully describe false vacuum decay, we expect that there should be one and only one negative mode. 

\subsection{Lee-Weinberg potential with \texorpdfstring{$D\geq 3$}{TEXT}}

The eigenequation reads (after rescaling)
\begin{equation}
\label{eq:eigenD}
\left[ -\partial^{2} - D \delta \left(r-1\right)  \right]  \Psi \left( \vec{x} \right)  =\lambda \Psi \left( \vec{x} \right)  
\end{equation}
We have suppressed the label for the eigenfunctions since we are only interested in the eigenvalues. To be specific, we are interested in negative modes and we shall take $\lambda=-\kappa^2$ with $\kappa>0$ below.

We  write $\Psi(\vec{x})=\sum_{l} \psi_{l}(r) Y_{l}(\vec{\theta})$ where $\Delta_{S^{D-1}} Y_{l}(\theta)=l(2-D-l) Y_{l}(\theta)$ with $\Delta_{S^{D-1}}$ being the Laplace-Beltrami operator. The eigenspectrum for fixed $l$ is degenerate with the degree of degeneracy for $D > 2$ given by\footnote{When analytically continuing this formula to noninteger $D$, ${\rm deg}(l;D)$ is noninteger except for $l=0$ when ${\rm deg}(0;D)=1$.}
\begin{align}
    {\rm deg}(l;D)=\frac{(D+l-3)!(D+2l-2)}{l!(D-2)!}\,.
\end{align}
For $D=2$, ${\rm deg}\left( l;2\right)  =2-\delta_{l0} $. Note also that a degeneracy of one only occurs for $l=0$ when $D\geq 2$. Consequently, the number of negative eigenvalues can be one, if and only if the negative eigenvalue occurs only for $l=0$. 

\begin{figure}[ht]
    \centering
    \includegraphics[scale=0.6]{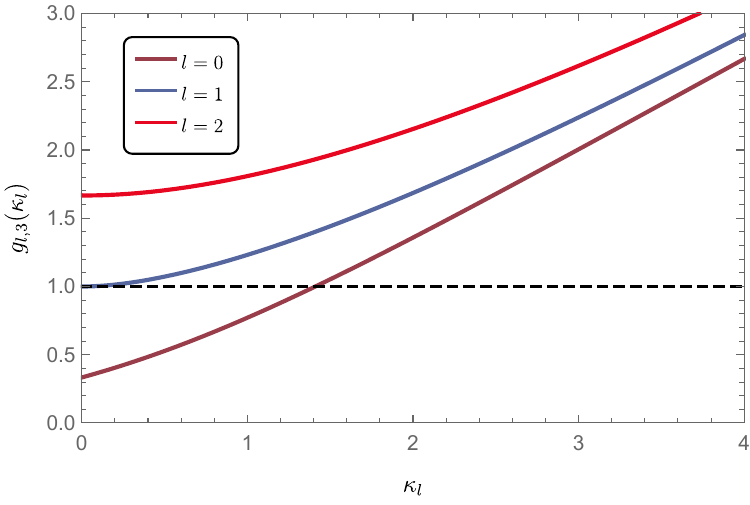}
    \includegraphics[scale=0.6]{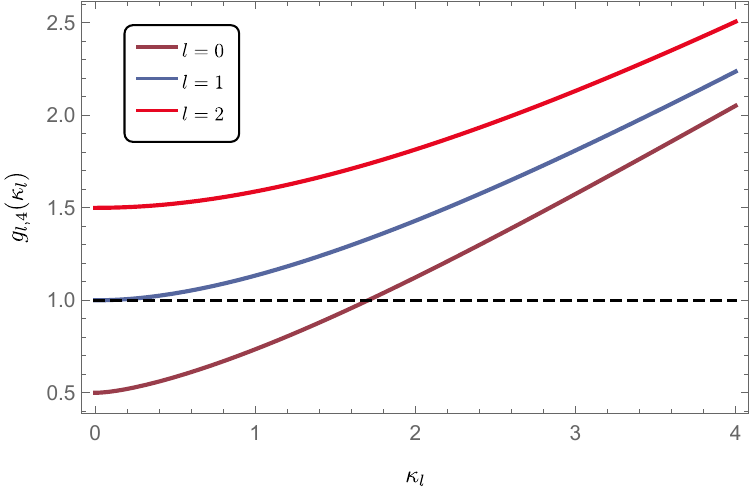}
    \caption{The function $g_{l,D}(\kappa_l)$ for the Lee-Weinberg potential for $D=3$ (left panel) and $D=4$ (right panel). A solution is indicated when the coloured curves intersect with the dashed line.}
    \label{fig:LW-g}
\end{figure}

Using the decomposition we arrive at
\begin{align}
    \left[ -\frac{1}{r^{D-1}} \frac{\d}{\d r} \left( r^{D-1}\frac{\d}{\d r} \psi_{l} \right)  +\frac{l\left( l+D-2\right)  }{r^{2}} \psi_{l} -D \delta \left( r-1\right)  \psi_l \right]  =-\kappa_l^2 \psi_l\,.
\end{align}
The solution of this equation is more evident in terms of 
\begin{align}
    \Phi_l(r)=r^{(D-2)/2}\psi_l(r)
\end{align}
together with a change of variable $\rho=\kappa_l r$. Then we get 
\begin{align}
\label{eq:besselradial}
\frac{1}{\rho}\frac{\d}{\d \rho} \left(\rho\frac{\d\Phi_{l}(\rho)}{\d \rho} \right)-\frac{1}{\rho^2}\left[\rho^2+ \nu_{l,D}^2 \right]  \Phi_{l}(\rho) = -\frac{D}{\kappa_l} \delta(\rho-\kappa_l)  \Phi_{l}(\rho)\,.
\end{align}
where
\begin{align}
    \nu_{l,D}^2=\left(l+\frac{D-2}{2} \right)^{2}\,.
\end{align}
Requiring $\Phi_l(0)=\Phi_l(\infty)=0$, we obtain the solution
\begin{align}
    \Phi_l(\rho)=\Theta(\kappa_l-\rho) c_1 I_{\nu_{l,D}} (\rho)+\Theta(\rho-\kappa_l) c_2 K_{\nu_{l,D}} (\rho)\,.
\end{align}
where $I_\nu$ and $K_\nu$ are modified Bessel functions of the first and third kind. Note that although Eq.~~\eqref{eq:besselradial} is the same for $\rho<\kappa_l$ and $\kappa_l<\rho$, the solutions are not the same because of the different boundary conditions.
The matching conditions at $\rho=\kappa_l$ are
\begin{subequations}
    \begin{align}
        &c_1 I_{\nu_{l,D}} (\kappa_l)=c_2 K_{\nu_{l,D}} (\kappa_l)\,, \label{eq:c1-c2}\\
        &c_2 K'_{\nu_{l,D}}(\kappa_l)-c_1 I'_{\nu_{l,D}} (\kappa_l) =-\frac{D}{\kappa_l} c_1 I_{\nu_{l,D}} (\kappa_l)\,. \label{eq:I-K-derivative}
    \end{align}
\end{subequations}
Substituting the first equation into the second gives the transcendental equation
\begin{align}
\label{eq:negative-mode-eq}
    g_{l,D}(\kappa_l)=1\,,
\end{align}
where
\be
g_{l,D}(\kappa_l)\equiv \frac{\kappa_l}{D}\left(\frac{I'_{\nu_{l,D}}(\kappa_l)}{I_{\nu_{l,D}} (\kappa_l)}- \frac{K'_{\nu_{l,D}}(\kappa_l)}{K_{\nu_{l,D}}(\kappa_l)}\right).
\ee
\cbl
The above equation can be solved numerically, showing a monotonically increasing behaviour of $g_{l,D}(x)$ with $x$. We plot $g_{l,D}(x)$ for $l=0,1,2$ and $D=3,4$ in Fig.~\ref{fig:LW-g}. It can be seen that there is a nonvanishing solution only for $l=0$. Since ${\rm deg}(0;D)=1$, this means that there is only one negative mode. From the curve for $l=1$, one can also see that there are $D$ zero modes in the sector $l=1$ as ${\rm deg}(1;D)=D$. The numerical results may be understood analytically on noting the asymptotic relations of the modified Bessel functions~\cite{Jeffrey2008-cl} and one can analytically confirm that there is only one solution for $l=0$ and $D$ vanishing solutions for $l=1$ and none for higher $l\geq 2$ for arbitrary $D$. Hence the bounce solution found here can be used to describe tunnelling in higher dimensional field theory as well.

\subsection{Triangular potential with \texorpdfstring{$D\geq 3$}{TEXT}}

For the triangular potential, let us first consider the case $D\geq 3$ as this case is similar to what has been discussed in the last subsection. The eigenequation reads
\begin{align}
    \left[-\partial^2- D\left(1+\frac{1}{c}\right)\delta(r-1)\right]\Psi(r)=-\kappa^2 \Psi(r)
\end{align}
where $r\in [0,\mathcal{R}]$ with $\mathcal{R}=(1+c)^{\frac{1}{D}}$. Following the previous procedure, we arrive at 
\begin{align}
\label{eq:besselradial2}
\frac{1}{\rho}\frac{\d}{\d \rho} \left(\rho\frac{\d\Phi_{l}(\rho)}{\d \rho} \right)-\frac{1}{\rho^2}\left[\rho^2+ \nu_{l,D}^2 \right]  \Phi_{l}(\rho) = -\frac{D (1+\frac{1}{c})}{\kappa_l} \delta(\rho-\kappa_l)\Phi_{l}(\rho)\,.
\end{align}
Requiring $\Phi_l(0)=0$, the general solution is
\begin{equation}
\label{eq:solution3}
\Phi_l(\rho)  =
\begin{cases}
    c_1 I_{\nu_{l,D}}(\rho)\,, \quad & 0<\rho < \kappa_l\,,   \\ c_{2} I_{\nu_{l,D}}(\rho) +c_{3}K_{\nu_{l,D}}(\rho)\,, \quad &\kappa_l<\rho< \kappa_l \mathcal{R}\,. 
\end{cases}   
\end{equation}
The matching conditions at $\rho =\kappa_l$ are
\begin{subequations}
\begin{align}
   & c_1  I_{\nu_{l,D}}(\kappa_l)  =c_2 I_{\nu_{l,D}}(\kappa_l)+ c_{3} K_{\nu_{l,D}}(\kappa_l)\,,\label{eq:matching2a} \\
   & c_2 I'_{\nu_{l,D}}(\kappa_l)  +c_3 K'_{{\nu_{l,D}}}(\kappa_l)-c_1 I'_{{\nu_{l,D}}}(\kappa_l) = -\frac{D\left(1+\frac{1}{c}\right)}{\kappa_l} c_1 I_{\nu_{l,D}}(\kappa_l)\label{eq:matching2b}\,. 
\end{align}   
\end{subequations}
The Dirichlet boundary condition at $\rho =\kappa_l\mathcal{R}$ gives
\begin{align}
   \label{eq:matching2c}
c_{2}I_{\nu_{l,D}}(\kappa_l \mathcal{R}) + c_{3}K_{{\nu_{l,D}}}(\kappa_l\mathcal{R})  =0\,. 
\end{align}
Combining all these equations, we find
\begin{align}
\label{eq:negative-mode-eq2}
h_{l,D}(\kappa_l)\equiv \left[1-\epsilon_{\nu_{l,D}} \frac{I_{\nu_{l,D}}\left(\kappa_l\right)}{K_{\nu_{l,D}}\left(\kappa_l\right)} \right]^{-1}  \frac{\kappa_l}{D(1+\frac{1}{c})}\left(\frac{I^{\prime }_{\nu_{l,D}}\left(\kappa_l \right)}{I_{\nu_{l,D}}\left(\kappa_l\right)} -\frac{K^{\prime}_{\nu_{l,D}}\left(\kappa_l \right)}{K_{\nu_{l,D}}(\kappa_l)}\right)=1\,,
\end{align}
where
\begin{align}
\epsilon_{\nu_{l,D}} \equiv\frac{K_{\nu_{l,D}}\left(\kappa_l\mathcal{R}\right)}{I_{\nu_{l,D}}(\kappa_l\mathcal{R})}\,. 
\end{align}
Note that in the limit $c\rightarrow\infty$, one has $\mathcal{R}\rightarrow\infty$, $\epsilon_{\nu_{l,D}}\rightarrow 0$ and thus that Eq.~\eqref{eq:negative-mode-eq2} coincides with Eq.~\eqref{eq:negative-mode-eq}. 

\begin{figure}
    \centering
    \includegraphics[scale=0.6]{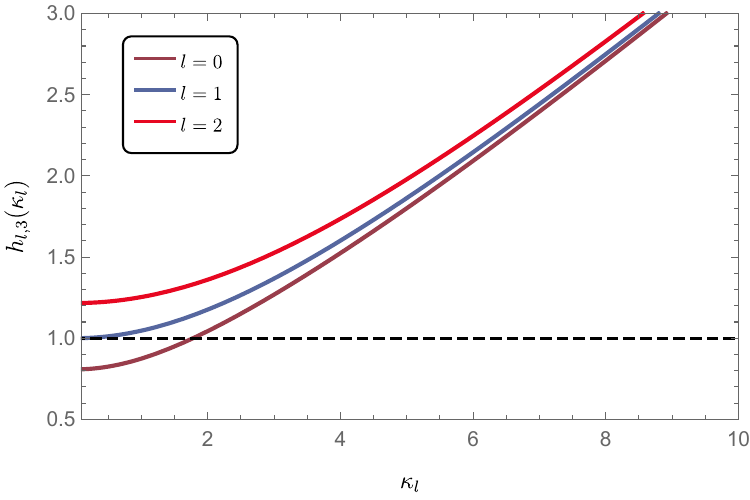}
    \includegraphics[scale=0.6]{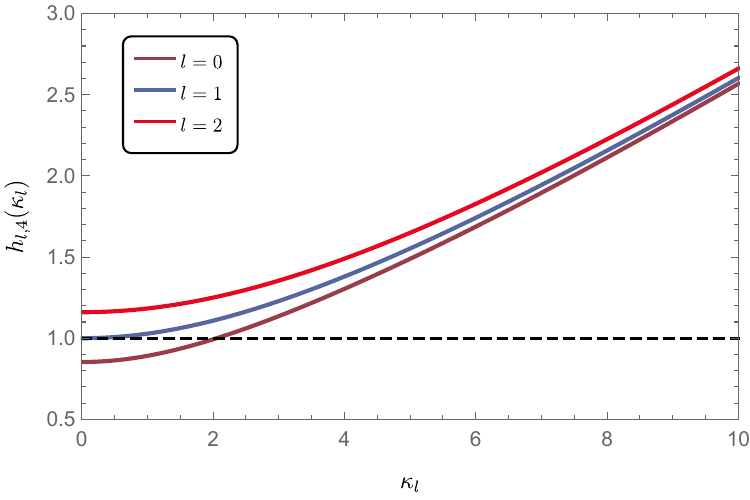}
    \caption{The function $h_{l,D}(\kappa_l)$ for the triangular potential for $D=3$ (left panel) and $D=4$ (right panel) and $c=1$. A solution is indicated when the coloured curves intersect the dashed line.}
    \label{fig:TriD3-D4}
\end{figure}

\begin{figure}[ht]
    \centering
    \includegraphics[scale=0.55]{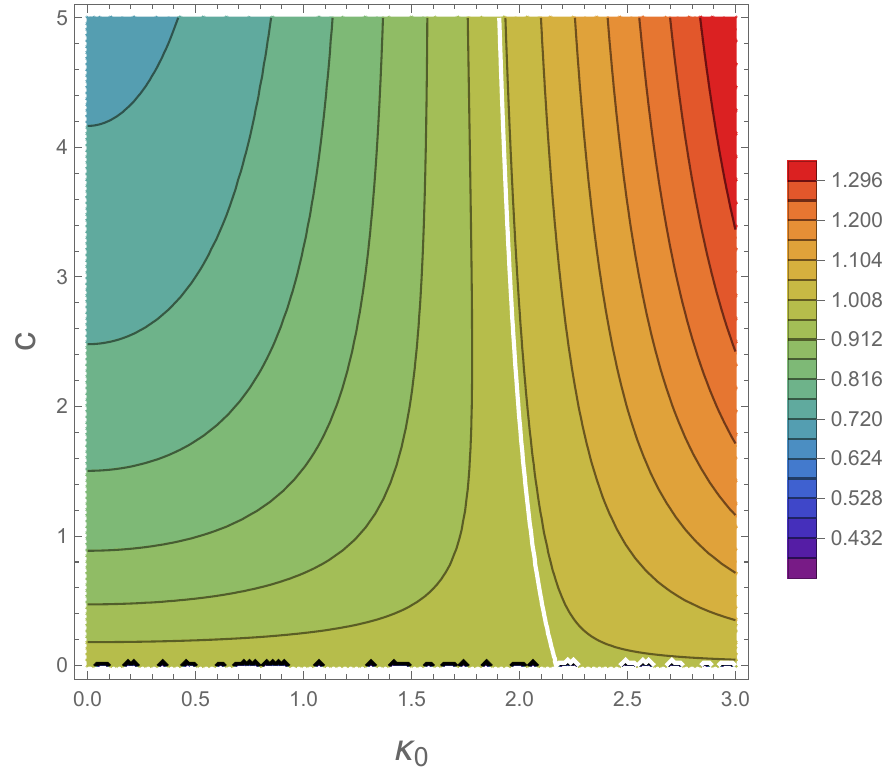}
    \caption{The function $h_{l,D}(\kappa_l)$ as a function of $\kappa_l$ and $c$ for $l=0$ and $D=4$. Eq.~\eqref{eq:negative-mode-eq2} is satisfied on the white thick line.}
    \label{fig:contour}
\end{figure}

Again, Eq.~\eqref{eq:negative-mode-eq2} can be solved numerically. We find that there is one solution only for $l=0$, which confirms that there is only one negative mode. In Fig.~\ref{fig:TriD3-D4}, taking  $c=1$, we plot $h_{l,D}(\kappa_l)$ for $l=0,1,2$, $D=3$ (left panel) and $D=4$ (right panel). To see how the parameter $c$ impacts the negative eigenvalue, we plot $h_{0,4}(\kappa_0)$ as a function of $\kappa_0$ and $c$ in Fig.~\ref{fig:contour}. It can be seen that the negative eigenvalue becomes smaller, i.e. $\kappa_0$ becomes larger (recall $\lambda=-\kappa_0^2$), for smaller $c$.

\cbl
\subsection{Triangular potential with \texorpdfstring{$D=1$}{TEXT}}

In this case, we have eigenequation
\begin{align}
    \left(-\partial_\tau^2-\left(1+\frac{1}{c}\right)\left[\delta(\tau+1)+\delta(\tau-1)\right]\right)\Psi(\tau)=-\kappa^2 \Psi(\tau)\,,
\end{align}
where $\tau\in (-\Upsilon,\Upsilon)$ with $\Upsilon=1+c$. The general solution is
\begin{align} 
\label{eq:eigenfunction-Tri-D1}
\Psi(\tau)=
\begin{cases}
    A_{1}\, \e^{\kappa \tau}+A_{2}\, \e^{-\kappa \tau }, \qquad &-\Upsilon< \tau <-1 \,,  \\
    G_1\,\e^{\kappa \tau}  +G_2\, \e^{-\kappa \tau},  \qquad  &-1< \tau < 1\,, \\
   F_{1}\, \e^{\kappa t}+F_{2}\, \e^{-\kappa \tau},  \qquad   &-1<\tau<\Upsilon\,.
\end{cases} 
\end{align}
The boundary conditions at $\tau=\pm \Upsilon$ gives
\begin{subequations}
\begin{align}
   &A_1\, \e^{-\kappa \Upsilon} + A_2 \, \e^{\kappa \Upsilon}=0\,,\\
   &F_1\, \e^{\kappa \Upsilon}  + F_2\, \e^{-\kappa \Upsilon} =0\,.
\end{align}
\end{subequations}
The matching conditions at $\tau=-1$ read
\begin{subequations}
    \begin{align}
      &G_1\,\e^{-\kappa} + G_2\,\e^{\kappa} =A_1\, \e^{-\kappa} + A_2 \, \e^{\kappa}\,,\\
      &G_1 \kappa\, \e^{-\kappa} - G_2\kappa\, \e^{\kappa} = A_1 \left[\kappa -\left(1+\frac{1}{c}\right) \right]  \e^{-\kappa} - A_2\left[\kappa +\left(1+\frac{1}{c}\right)\right]  \e^{\kappa}\,,
    \end{align}
\end{subequations}
while at $\tau=1$ are
\begin{subequations}
\begin{align}
   &F_1\,\e^{\kappa}+ F_2\,\e^{-\kappa} = G_1\,\e^{\kappa} + G_2\,\e^{-\kappa}\,,\\
   &F_1\kappa\, \e^{\kappa}-F_2\kappa\, \e^{-\kappa} =
   G_1\left[\kappa -\left(1+\frac{1}{c}\right) \right] \e^{\kappa} -G_2\left[\kappa +\left(1+\frac{1}{c}\right) \right]  \e^{-\kappa}\,.
\end{align}
\end{subequations}
Combing all the above equations, we obtain
\begin{align}
\label{eq:negative-mode-eq3}
f(\kappa)\equiv f_{0}(\kappa)+f_{1}(\kappa)\epsilon(\kappa) +f_{2}(\kappa)\epsilon(\kappa)^{2} =0
\end{align}
where $\epsilon=\exp(-2 \kappa \Upsilon)$ and 
\begin{subequations}
\begin{align}
f_{0}&=-\alpha^{2} +\e^{4\kappa}  \left(2\kappa-\alpha \right)^{2}\,,  \\
f_{1}&=2\alpha\,\e^{2\kappa} \left[ \alpha +2\kappa +\left( 2\kappa -\alpha \right)  \e^{4\kappa}  \right]\,,  \\
f_{2}&=\e^{8\kappa} \alpha^{2} -\e^{4\kappa}   \left( \alpha +2\kappa \right)^{2}\,,
\end{align}
\end{subequations}
with $\alpha\equiv 1+1/c>1$. Note that $\epsilon$ is a function of $\kappa$.

\begin{figure}[ht]
    \centering
    \includegraphics[scale=0.55]{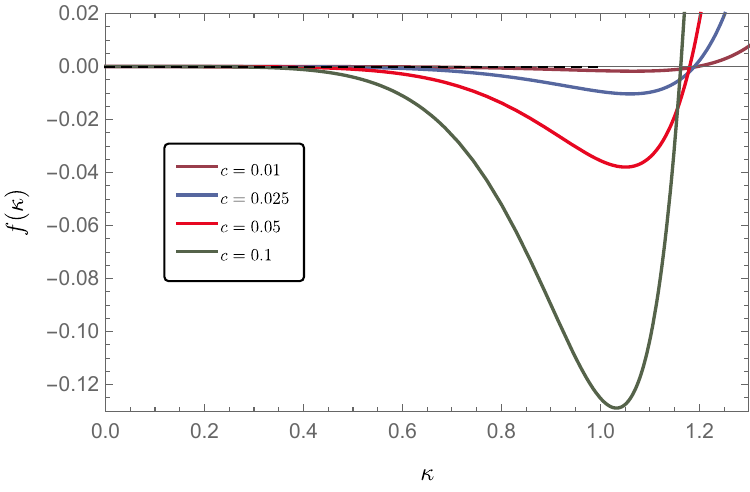}
    \includegraphics[scale=0.55]{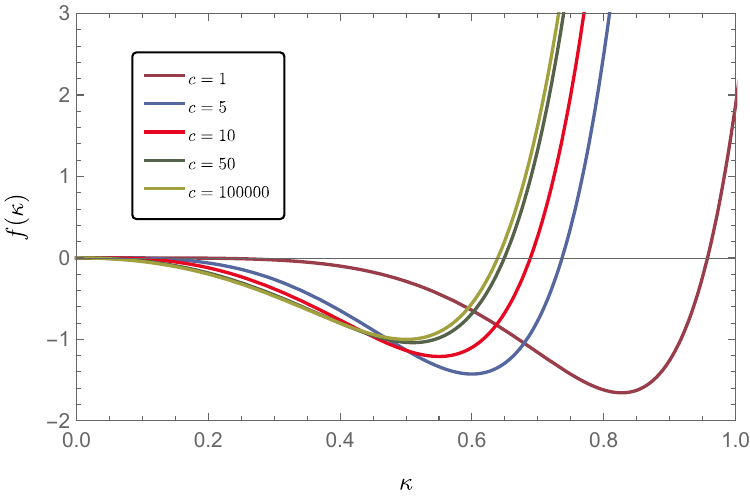}
    \caption{$f(\kappa)$ defined in Eq.~\eqref{eq:negative-mode-eq3} for small values of $c$ (left panel) and bigger values of $c$ (right panel).
    The different curves intersect the axis $f(\kappa)=0$ only once away from $\kappa=0$.}
    \label{fig:fk}
\end{figure}

We solve Eq.~\eqref{eq:negative-mode-eq3} numerically and find that there is one and only one positive solution for $c\in (0,\infty)$. We plot $f(\kappa)$ in Fig.~\ref{fig:fk} for relatively smaller values of $c$ (left panel), and bigger values of $c$ (right panel), respectively. It can be seen that one has a larger positive root $\kappa$ (which gives a smaller negative eigenvalue $-\kappa^2$) for a smaller value of $c$, in agreement with the higher-dimensional cases. The value of the eigenvalue approaches a limit quickly when $c$ increases.

\section{Conclusion and discussion}
\label{sec:conclusion}

In this paper, we have discussed the piecewise linear potentials, specifically the Lee-Weinberg potential introduced in Ref.~\cite{Lee:1985uv} and the triangular-shaped potential introduced in Ref.~\cite{Duncan:1992ai}, as simplified false vacuum decay models. We calculate the bounce solutions for general spacetime dimension $D$ and for $D=4$ results in the literature are recovered. A condition for such analytic bounce solutions to describe tunnelling is that there should be an odd number of negative modes. We show that there is one and only one negative mode in the spectrum of the fluctuation operator at the bounce for arbitrary $D$. This is achieved by explicitly solving the eigenequation for the negative modes. An advantage of such piecewise linear potentials is that the eigenequations take a very simple Schr\"{o}dinger-like equation with delta potentials in a universal form. 

\cb
We have concentrated on potentials with one or two discontinuities in $U''(\phi)$. The general case is cumbersome but considering a variant of our case involving a harmonic potential around the false vacuum is one way of checking the robustness of our conclusions. This analysis is given in appendix \ref{app:tri-mass} and our conclusions remain unchanged. A harmonic potential can, after all, be approximated by many straight-line sectors, which will lead to many delta function terms in the equation for the spectrum of fluctuations. 
\cbl

Piecewise linear potentials considered here could have several applications. For example, the triangular potential was used in studying metastable supersymmetric minima in Refs.~\cite{Intriligator:2006dd,Carena:2012mw}, dark energy~\cite{Pastras:2011zr} and gravitational wave production from cosmological first-order phase transitions~\cite{Jaeckel:2016jlh}. The unbounded-from-below Lee-Weinberg potential is similar to those that appear in theories with \cPT~symmetry~\cite{Bender:2015uxa,Bender:2018pbv,Bender:2021fxa,Felski:2021evi}. Potentials unbounded from below lead to instability of the vacuum in the conventional Hermitian framework. In the presence of \cPT~symmetry, there is a ``phase''~\cite{Mavromatos:2024ozk} without the instability, known as the~\cPT-symmetric phase~\cite{Bender:2007nj}. For $D=1$, using a WKB treatment in a quantum mechanical framework, it is known that the eigenvalues for the Hamiltonian are real~\cite{Bender:1998ke,R2}. It is believed that higher dimensional generalisations of this result also hold. A way to investigate this for $D > 1$ is in terms of Euclidean path integrals and fluctuations around bounce-like solutions~\cite{Ai:2022csx} through the approach here. When extending the present analysis to polygonal potentials with more segments, the analysis may also justify the validity of using the latter for efficient computation of bubble nucleation rates~\cite{Guada:2018jek} (cf. other recent works~\cite{Ekstedt:2023sqc,Bai:2024pii}).

\vskip 2cm

\section*{Acknowledgments}
The authors of this work are supported by the Engineering and Physical Sciences Research Council (grant No. EP/V002821/1). JA is also supported by the Leverhulme Trust (grant No. RPG-2021-299) and by the Science and Technology Facilities Council (grant No.STFC-ST/X000753/1). 

\begin{appendix}

\section{\cb Colemans's conjecture}
\label{app:Colemann}

\cb We review here Coleman's argument as to why there should be one and only one negative mode in the spectrum of the fluctuations about the bounce. It is intuitively reasonable but, as stated by Coleman himself, it is not based on functional methods and so the extension to field theory is not rigorously proven. 
Colemann's argument is based on a finite set of coupled oscillators. 
Independently of the work of Coleman, the study of tunnelling in systems of sufficiently strongly interacting oscillators has been done in much detail, over many years, within the context of molecular physics and Bose-Einstein condensates \cite{Keshavamurthy2024-my}. While tunnelling in quantum field theory is dramatically simplified with the assumed $O(D)$ symmetry for the bounce solution, it is not clear how a similar symmetry can be imposed for coupled oscillators. Moreover the chaotic behaviour for the classical trajectories of the oscillators might introduce complications~\cite{Escande1985-yo}.

The dynamics of $N$ oscillators, which can be collectively identified with a vector $\vec q =\left(q^{1},\ldots,q^{N} \right)$, is governed by the (Euclidean) Lagrangian 
\begin{align}
\label{nonint}
 L=\frac{1}{2} \sum_{j=1}^{N} \left( \frac{\d q^{j}}{\d t} \right)^{2} +V\left(q^{1},q^{2},\ldots ,q^{N} \right) = \frac{1}{2} \left( \frac{\d\vec{q}}{\d t} \right)^{2} +V\left( \vec{q} \right).
\end{align}
where $V\left(\vec{q} \right)$ is the potential energy of the oscillators. Without loss of generality, one can choose the false vacuum to be the origin $\vec{q}=0$ and $V(0)=0$. 
$\vec{q}=0$ is thus a local minimum of the potential, surrounded by a region with positive $V$. Since the potential allows for tunnelling, there must be regions with $V<0$ far away from the origin. Then $V=0$ defines a surface (denoted as $\Sigma$) that separates the near region with positive $V$ and the further region with negative $V$. Barrier penetration is characterised by a trajectory $\bar{q}(l)$ that starts from $\vec{q}=0$ and ends at a point on $\Sigma$ that minimise 
\begin{align}
    S_{\rm B}[\vec{q}(l)]=\int \d l \, \sqrt{2 V(\vec{q}(l))}\,,
\end{align}
where $l$ parameterises the trajectory and
\begin{align}
    \d l^2= \delta_{ij}\d q^i \d q^j\,.
\end{align}
The subscript ``B'' in $S_{\rm B}$ denotes barrier penetration. $\bar{q}(l)$ is called the most probable escape path (MPEP)~\cite{PhysRevD.8.3346}. The coefficient $B$ in Eq.~\eqref{eq:decay_rate} is then given by $B=2 S_{\rm B}[\bar{q}(l)]$. 
Since $\bar{q}(l)$ minimise $S_{\rm B}[\vec{q}(l)]$, we have 
\begin{align}
    \left.\delta^2 S_{\rm B}\right|_{\bar{q}(l)} \geq 0\,.
\end{align}
This condition was assumed, but as noted by Coleman \cite{Coleman:1987rm}, cannot be rigorously justified since the set of admissible paths is not an open set. On the other hand, we have the Euclidean action 
\be 
S'_{\rm E}[ \vec q ] =\int_{\tau_i}^{\tau_f} \d\tau \left[ \frac{1}{2} \left( \frac{\d\vec q}{\d\tau} \right)^{2}  +\, V\left(\vec q \right) \right]
\ee
If we consider motions that start from $\vec{q}=0$ and end at a point on $\Sigma$ within a fixed time interval ($\tau_f-\tau_i$), then there is a one-to-one map between a $\vec{q}(\tau)$ that minimises $S'_{\rm E}[\vec{q}(\tau)]$ and a $\vec{q}(l)$ that minimises $S_{\rm B}[\vec{q}(l)]$, and 
\begin{align}
    S'_{\rm E}[\vec{q}(\tau)]=S_{\rm B}[\vec{q}(l)]\,.
\end{align}
Corresponding to the MPEP $\bar{\vec{q}}(l)$, there is thus a unique Euclidean motion $\bar{\vec{q}}(\tau)$, for a given time interval, that satisfies the Euler-Lagrange equation  
$$\frac{\d^{2}q_{j}}{\d\tau^{2}} =\frac{\partial V}{\partial q_{j}}\,.$$
And because of the correspondence, we have 
\begin{align}
\label{eq:56}
    \left.\delta^2 S'_{\rm E}\right|_{\bar{\vec{q}}(\tau)}\geq 0\,.
\end{align}

For tunnelling, $\bar{\vec{q}}(\tau)$ needs to have vanishing velocity $\d \bar{\vec{q}}/d\tau$ at the initial time. Therefore, $\bar{\vec{q}}(\tau)$ is a zero-energy solution and it approaches the origin (false vacuum) in infinite time. So we can consider a $\bar{\vec{q}}(\tau)$ with $\tau$ occupying the range $(-\infty,0]$. The key observation is that the bounce solution can be constructed from $\bar{\vec{q}}(\tau)$. By energy conservation, $\left.\d \bar{\vec{q}}/\d \tau\right|_{\tau=0} =0$. Then one can extend $\bar{\vec{q}}(\tau)$ to the whole range $(-\infty,\infty)$ by reflection $\bar{\vec{q}}(-\tau)=\bar{\vec{q}}(\tau)$. We denote the extended solution as $\vec{q}_b(\tau)$. This extended solution obviously satisfies the Euler-Lagrange equation and satisfies the boundary conditions of the bounce: $\vec{q}_b(\pm \infty)=0$, $\d \vec{q}_b/\d\tau|_{\tau=0}=0$. Therefore, $\vec{q}_b(\tau)$ is just the bounce.

The extended solution stationarises the usual Euclidean action
\begin{align}
   S_{\rm E}[\vec q ] =\int_{-\infty}^{\infty} \d\tau \left[ \frac{1}{2} \left( \frac{\d\vec q}{\d\tau} \right)^{2}  +\, V\left(\vec q \right) \right]\,.
\end{align}
Now let us consider small variations $\delta \vec{q}(\tau)$ about $\vec{q}_b(\tau)$. If we fix $\delta \vec{q}(0)=0$, then the variations in the range $(-\infty, 0]$ and $[0,\infty)$ do not spoil the boundary conditions defined for $S'_{\rm E}$ (when taking $\tau_i=-\infty,\tau_f=0$ and the reflected version). In this case we have 
\begin{align}
    \delta S_{\rm E}=2 \delta S'_{\rm E}\,.
\end{align}
By Eq.~\eqref{eq:56}, we thus have $\delta S_{\rm E}\geq 0$ for fluctuations with $\delta \vec{q}(0)=0$. Therefore, negative eigenmodes in the spectrum of $\delta^2 S_{\rm E}$ must give fluctuations that violate $\delta \vec{q}(0)= 0$.

By the well-known node theorem and the fact that $\dot{q}_b$ (which is an odd function in $\tau$) is a zero mode, we know there must be at least one negative mode since the lowest mode cannot be odd. However, if there are two or more negative modes, then again we can consider the superposition of the fluctuations corresponding to these negative modes, which should cause $\delta S_{\rm E}<0$, such that $\delta q(0)=0$. (With only one negative mode, one cannot do this). However, we have shown that once $\delta q(0)=0$, $\delta S_{\rm E}$ cannot be negative and this is a contradiction. Therefore, there can be one and only negative mode. This is the essence of Coleman's argument. For more details, see Ref.~\cite{Coleman:1987rm}.

\cbl
\section{\texorpdfstring{$D=2$}{TEXT} for the triangular potential}
\label{app:tri-D2}

Unlike the case of the Lee-Weinberg potential, a bounce solution for the triangular potential for $D=2$ exists. The bounce solution is found to be\footnote{We have assumed that $\phi_*\leq \phi_-$, which should impose a condition for the parameters via Eq.~\eqref{eq:phistar}.}
\begin{align}
    \label{eq:bounce-tri-D2}
    \phi_b(r)=\begin{cases} 
    -\frac{\lambda_- r^2}{4}+\phi_*\,, \qquad &{\rm for \ } 0\leq r< r_1\,, \\
    \frac{\lambda_+ r^2}{4}-2 \left[\left(\phi_*-\phi_T\right)\left(1+\frac{1}{c}\right)\right]\log\left(\frac{r}{R}\right)-\left[\left(\phi_*-\phi_T\right)\left(1+\frac{1}{c}\right)\right]\,, &{\rm for\ } r_1\leq r<R\\
    0\,, \qquad &{\rm for \ } r\geq R\,,
    \end{cases}
\end{align}
where 
\begin{align}
r_1=2\sqrt{\frac{\phi_*-\phi_T}{\lambda_-}}\,,  \quad R=2\sqrt{\frac{\left(\phi_*-\phi_T\right)\left(1+\frac{1}{c}\right)}{\lambda_+}}\,,
\end{align}
and $\phi_*$ is determined implicitly by the matching conditions at $r=r_1$, leading to
\begin{align}
\label{eq:phistar}
    \left(1+\frac{\lambda_+}{\lambda_-}\right)\log\left(\frac{4(\phi_*-\phi_T)}{\lambda_-}\right)+\frac{\phi_*}{\phi_*-\phi_T}=4\left(1+\frac{\lambda_+}{\lambda_-}\right)\sqrt{(\phi_*-\phi_T)\left(\frac{1}{\lambda_+}+\frac{1}{\lambda_-}\right)}\,.
\end{align}


\section{\cb Adding a mass term to the triangular potential}
\label{app:tri-mass}

\cb 
For the triangular potential considered in the main text, the field reaches the false vacuum at finite time or finite Euclidean radius in the bounce solution. This is because the potential is not smooth at the false vacuum. One can avoid this by replacing the linear function with a mass term at $\phi=0$ (false vacuum).\footnote{We thank the anonymous referee for suggesting this option.} In this section, we study such a modification and consider the potential
\begin{align}
\label{eq:U-Tri-mass}
    U(\phi)=\begin{cases}
   \frac{1}{2}m^2\phi^2\,, \quad &{\rm for \ } 0<\phi<\phi_T\,, \\
    V_T-\lambda (\phi-\phi_T)\,, \quad &{\rm for\ } \phi_T<\phi<\phi_-\,,
    \end{cases}
\end{align}
where continuity requires $V_T=m^2\phi_T^2/2$. The first derivative is
\bea
U'(\phi)=\begin{cases}
    m^2\phi \quad & \mbox{for}\ 0\leq\phi<\phi_T\,,\\
    -\lambda \  & \mbox{for}\ \phi_T\leq\phi\leq\phi_-\,,
\end{cases}
\eea
such that
\be
U''(\phi)=m^2\Theta(\phi_T-\phi)-(m^2\phi_T+\lambda)\delta(\phi-\phi_T)\,.
\ee

\subsection{\texorpdfstring{$D=1$}{TEXT} }
The turning point $\phi_\ast$ for the bounce $\phi_b$ can be found from energy conservation
\be
\frac{1}{2}(\dot\phi)^2- U(\phi)=~0~=- U(\phi_\ast)\quad \Rightarrow \quad \phi_\ast=\phi_T+V_T/\lambda\,.
\ee
We define the origin of time such that $\phi_b(0)=\phi_\ast$ for the bounce to be an even function of time. 
We also define the time $\tau_1$ such that $\phi_b(\pm\tau_1)=\phi_T$. Then the bounce solution is found to be
\begin{align}
\label{eq:bounce-Tri-mass-D1}
  \phi_b(\tau)=\begin{cases} 
    \phi_T\,\e^{m(\tau+\tau_1)} \,, \qquad &{\rm for \ } -\infty< \tau<-\tau_1\,, \\
    -\frac{1}{2}\lambda \tau^2+\phi_* \,, \qquad &{\rm for\ } -\tau_1 \leq \tau \leq\tau_1\,,\\
    \phi_T\, \e^{m(-\tau+\tau_1)}\,, \qquad &{\rm for \ }  \tau_1<\tau<\infty \,,
    \end{cases}
\end{align}
The value for $\tau_1$ is found by imposing the continuity of the bounce, $\lambda\tau_1=\sqrt{2V_T}$.
The bounce action is then
\be
B=2V_T\left(\frac{1}{m}+\frac{2\sqrt{2V_T}}{3\lambda}\right)\,.
\ee

For the fluctuation operator, we note that
\be
\delta(\phi_b-\phi_T)=\frac{\delta(\tau+\tau_1)}{|\dot\phi_b(-\tau_1)|}+\frac{\delta(\tau-\tau_1)}{|\dot\phi_b(\tau_1)|}
=\frac{1}{\sqrt{2V_T}}\Big(\delta(\tau+\tau_1)+\delta(\tau-\tau_1)\Big)\,,
\ee
such that
\be
U''(\phi_b)=m^2\Theta(\phi_T-\phi_b)-\alpha\Big(\delta(\tau+\tau_1)+\delta(\tau-\tau_1)\Big)\,,
\ee
where 
\be
\alpha\equiv\frac{m^2\phi_T+\lambda}{\sqrt{2V_T}}=m+\frac{1}{\tau_1}\,.
\ee

\begin{figure}
    \centering   \includegraphics[width=0.5\linewidth]{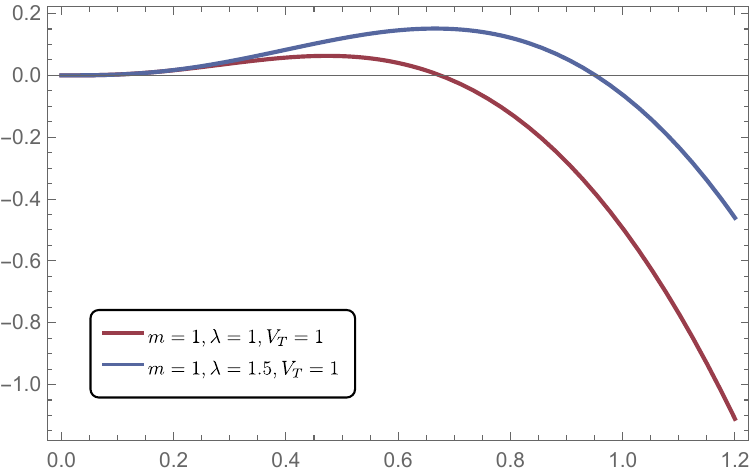}
    \caption{$f(\kappa)$ defined in Eq.~\eqref{eq:fkappa} for two choices of parameters (in arbitrary units). $f(\kappa)=0$ gives the zero (when $\kappa=0$) and negative modes.}
    \label{fig:mass-tri-D1}
\end{figure}

The eigenvalue equation (for negative values) to solve is
\be
\Big(-\partial_\tau^2+U''(\phi_b)\Big)\Psi=-\kappa^2\Psi\, ,
\ee
where as before we assume $\kappa>0$. 
Defining $\omega=\sqrt{m^2+\kappa^2}$, we then have
\begin{align} 
\label{eq:eigenfunction-Tri-mass-D1}
\Psi(\tau)=
\begin{cases}
    A\, \e^{\omega \tau}\,, \qquad &-\infty < \tau <-\tau_1 \,,  \\
    G_1\,\e^{\kappa \tau}  +G_2\, \e^{-\kappa \tau},  \qquad  &-\tau_1< \tau < \tau_1\,, \\
    F\, \e^{-\omega \tau},  \qquad   &\tau_1<\tau<\infty \,.
\end{cases} 
\end{align}
The matching conditions require $\Psi$ to be continuous at $\pm\tau_1$. Also, an integration of the eigenvalue equation
between $\pm\tau_1-\epsilon$ and $\pm\tau_1+\epsilon$ leads, in the limit $\epsilon\to0$, to the discontinuity $-\alpha\Psi$ in $\dot\Psi$.
At $\tau=-\tau_1$ we have then
\begin{subequations}
\bea   
G_1\,\e^{-\kappa\tau_1}+G_2~\e^{\kappa\tau_1}&=&A\,\e^{-\omega\tau_1}\,,\\
G_1\kappa\,\e^{-\kappa\tau_1}-G_2\kappa\,\e^{\kappa\tau_1}&=&A(\omega-\alpha)\,\e^{-\omega\tau_1}\,.
\eea
\end{subequations}
At $\tau=\tau_1$ we have
\begin{subequations}
\bea
G_1\,\e^{+\kappa\tau_1}+G_2\,\e^{-\kappa\tau_1}&=&F\,\e^{-\omega\tau_1}\,,\\
G_1(\kappa-\alpha)\e^{+\kappa\tau_1}-G_2(\kappa +\alpha)\e^{-\kappa\tau_1}&=&-F\omega\,\e^{-\omega\tau_1}\,.
\eea
\end{subequations}
Rearranging these equations, we are left with a homogeneous linear system for $G_1,G_2$, whose vanishing determinant leads to
\be
\label{eq:fkappa}
f(\kappa)\equiv (\omega-\kappa-\alpha)^2 \e^{-4\kappa\tau_1}-(\omega+\kappa-\alpha)^2 =0\,.
\ee 
One can easily see that $\kappa=0$ is a solution, corresponding to the expected zero mode of the fluctuation operator. We numerically checked that there is one and only one negative mode. Two examples are shown in Fig.~\ref{fig:mass-tri-D1}.

\subsection{\texorpdfstring{$D\geq 2$}{TEXT}}

\begin{figure}[ht]
    \centering
    \includegraphics[width=0.5\linewidth]{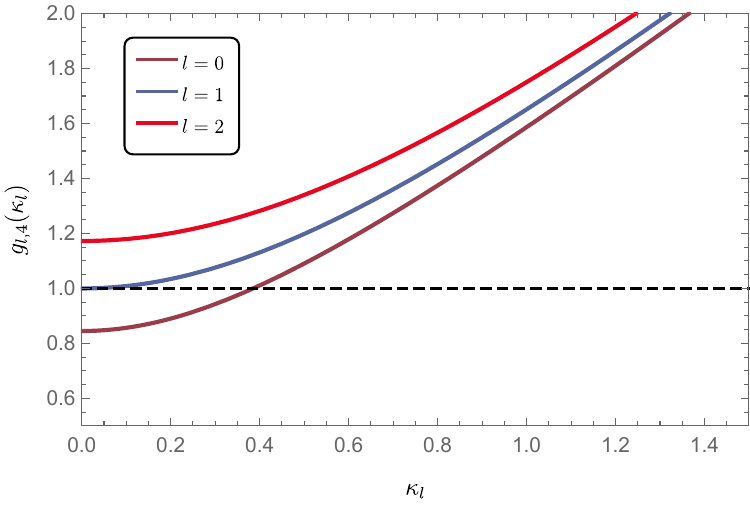}
    \caption{An numerical example of the function $g_{l,D}(\kappa_l)$ given in Eq.~\eqref{eq:glD-tri-mass}. Here we have chosen $m=1$, $\lambda=1$, $\phi_T=1$ (all with arbitrary units) and $D=4$.}
    \label{fig:tri-mass}
\end{figure}

Now we extend the above analysis to $D\geq 2$. Again, we define $r_1$ by $\phi_b(r_1)=\phi_T$. Then the bounce is found to be
\begin{align}
\label{eq:bounce-tri-msdd-Dgeq2}
    \phi_b(r)=\begin{cases} 
    \phi_{b,1}(r)\equiv -\frac{\lambda}{2D}r^2+\phi_*\,, \qquad &{\rm for \ } 0\leq r< r_1\,, \\
     \phi_{b,2}(r)\equiv\phi_T\left(\frac{r}{r_1}\right)^{\frac{2-D}{2}}\frac{K_\nu(mr)}{K_\nu(m r_1)}\,,\qquad &{\rm for\ } r_1\leq r< \infty\,,
    \end{cases}
\end{align}
where 
\begin{align}
    \nu=\frac{D-2}{2}\,,\qquad r_1=\left(\frac{2D(\phi_*-\phi_T)}{\lambda}\right)^{\frac{1}{2}}\,.
\end{align}
Above, $\phi_*$ is implicitly determined by the matching condition
\begin{align}
    \left.\frac{\d \phi_{b,1}(r)}{\d r}\right|_{r=r_1}=\left.\frac{\d\phi_{b,2}(r)}{\d r}\right|_{r=r_1}\,.
\end{align}
For the eigenequations, the same procedure as before leads to
\begin{align}
\frac{1}{\rho}\frac{\d}{\d \rho} \left(\rho\frac{\d\Phi_{l}(\rho)}{\d \rho} \right)-\frac{1}{\rho^2}\left[\rho^2+ \nu_{l,D}^2 +\frac{\rho^2 m^2}{\kappa_l^2} \Theta(r-r_1)\right]  \Phi_{l}(\rho) = -\frac{(m^2\phi_T+\lambda )}{|\phi'_b(r_1)|\kappa_l} \delta(\rho-\kappa_lr_1)\Phi_{l}(\rho)\,,
\end{align}
where 
\begin{align}
    \nu_{l,D}=\left(l+\frac{D-2}{2}\right)^2\,.
\end{align}
The general solution with $\Phi_l(0)=\Phi_l(\infty)=0$ is
\begin{align}
    \Phi_l(\rho)=\Theta(\kappa_l r_1-\rho) c_1 I_{\nu_{l,D}} (\rho)+\Theta(\rho-\kappa_l r_1) c_2 K_{\nu_{l,D}} \left(\frac{\omega_l}{\kappa_l}\rho\right)\,,
\end{align}
where $\omega_l=\sqrt{m^2+\kappa_l^2}$.
Again, imposing the continuity of the $\Phi_l(\rho)$ and its derivative at $\kappa_l r_1$ finally gives 
\begin{align}
    g_{l,D}(\kappa_l)=1
\end{align}
with
\begin{align}
\label{eq:glD-tri-mass}
    g_{l,D}(\kappa_l)=\frac{|\phi'_b(r_1)|\kappa_l}{m^2 \phi_T +\lambda}\left[\frac{I'_{\nu_{l,D}}(\kappa_l r_1)}{I_{\nu_{l,D}}(\kappa_l r_1)}-\frac{K'_{\nu_{l,D}}(\omega_l r_1)}{K_{\nu_{l,D}}(\omega_l r_1)}\frac{\omega_l}{\kappa_l}\right]\,.
\end{align}
We show $g_{l,D}(\kappa_l)$ with an example choice of parameters in Fig.~\ref{fig:tri-mass} and confirm that indeed there is one and only one negative mode (in the $l=0$ sector). We also see that there is a zero mode with $l=1$. The situation is very similar to that of the Lee-Weinberg potential, shown in Fig.~\ref{fig:LW-g}.

\cbl
\end{appendix}

\bibliographystyle{utphys}
\bibliography{bibliLP}

\end{document}